\documentclass[aps, prd, twocolumn, superscriptaddress, nofootinbib,
amsmath, amssymb,floatfix]{revtex4-2}

\usepackage{graphicx}
\usepackage{mathrsfs,multirow} 
\newcommand{\mlight}{m_{\rm ud}}
\newcommand{\rmt}{{\rm RMT}}

\begin{document}

\title{Localization properties of Dirac modes at the Roberge-Weiss
  phase transition}

\author{Marco Cardinali}
\email{marco.cardinali@pi.infn.it}
\affiliation{Dipartimento di Fisica dell'Universit\`a di Pisa and INFN
  - Sezione di Pisa, Largo Pontecorvo 3, I-56127 Pisa, Italy}

\author{Massimo D'Elia}
\email{massimo.delia@unipi.it}
\affiliation{Dipartimento di Fisica dell'Universit\`a di Pisa and INFN
  - Sezione di Pisa, Largo Pontecorvo 3, I-56127 Pisa, Italy}

\author{Francesco Garosi}
\email{fgarosi@sissa.it}
\affiliation{SISSA, Via Bonomea 265, 34136, Trieste, Italy}

\author{Matteo Giordano}
\email{giordano@bodri.elte.hu}
\affiliation{ELTE E\"otv\"os Lor\'and University, Institute for
  Theoretical Physics, P\'azm\'any P\'eter s\'et\'any 1/A, H-1117, Budapest,
  Hungary}

\begin{abstract}
  We study the localization properties of the low-lying Dirac
  eigenmodes in QCD at imaginary chemical potential $\hat{\mu}_I=\pi$
  at temperatures above the Roberge-Weiss transition temperature
  $T_{\rm RW}$. We find that modes are localized up to a
  temperature-dependent ``mobility edge'' and delocalized above it,
  and that the mobility edge extrapolates to zero at a temperature
  compatible with $T_{\rm RW}$. This supports the existence of a
  strong connection between localization of the low Dirac modes and
  deconfinement, studied here for the first time in a model with a
  genuine deconfinement transition in the continuum limit in the
  presence of dynamical fermions.
\end{abstract}

\maketitle

\section{Introduction}  

The interest in gauge theories at nonzero imaginary chemical potential
is due both to practical and theoretical reasons. On the one hand,
they are a means to side-step the notorious sign problem encountered
at real chemical potential: being free from the sign problem, they
allow direct numerical simulations with importance sampling methods,
from which one can attempt an analytic continuation to the physically
relevant case of real chemical
potential~\cite{deForcrand:2002hgr,DElia:2002tig,DElia:2009pdy,
  Cea:2006yd,Cea:2014xva,Bonati:2014kpa,Cea:2015cya,Bonati:2015bha,
  Bellwied:2015rza,DElia:2016jqh,Gunther:2016vcp,Alba:2017mqu,
  Vovchenko:2017xad,Bonati:2018nut,Borsanyi:2018grb,Bellwied:2019pxh,
  Borsanyi:2020fev}.  On the other hand, they provide an interesting
testing ground to study the interplay of dynamical fermions and the
center symmetry of the pure gauge theory, and how this affects the
phase diagram of the theory.

As is well known, the analogue of center symmetry in the case of
SU($N_c$) theories with dynamical fundamental fermions is the
Roberge-Weiss symmetry~\cite{Roberge:1986mm}, that states that the
partition function is periodic in the reduced imaginary chemical
potential $\hat{\mu}_I = \mu_I/T$ with period $2\pi/N_c$. This
periodicity is realized differently at low and at high temperatures:
while analytic in $\hat{\mu}_I$ at low $T$, at high $T$ the partition
function displays lines of first order phase transitions at
$\hat{\mu}_I=(2k+1)\pi/N_c$.  These transitions correspond to a change
in the center sector favored by the fermions.  The RW transition lines
and their endpoints have been thoroughly investigated by lattice
simulations~\cite{deForcrand:2002hgr,DElia:2002tig, DElia:2007bkz,
  Cea:2009ba, DElia:2009bzj,Bonati:2010gi,
  deForcrand:2010he,Cea:2012ev, Philipsen:2014rpa,Alexandru:2013uaa,
  Wu:2013bfa,Wu:2014lsa,Wu:2016vai,
  Bonati:2014kpa,Nagata:2014fra,Makiyama:2015uwa,
  Czaban:2015sas,Pinke:2015ylw,Kashiwa:2016vrl,Kashiwa:2015tna,
  Bonati:2016pwz,Bonati:2018fvg} and effective
models~\cite{Kouno:2009bm,Sakai:2009dv,Sakai:2010rp,
  Sasaki:2011wu,Kouno:2011zu,Aarts:2010ky,Rafferty:2011hd,
  Morita:2011eu,Kashiwa:2011td,Pagura:2011rt,Scheffler:2011te,
  Kashiwa:2013rm,Kashiwa:2012xm}. Depending on the value of the quark
masses, the first order lines end at a second-order point at
$T_{\rm RW}$, or alternatively at a triple point, which is connected
via a (pseudo)critical line to the (pseudo)critical temperature at
vanishing chemical potential.

A growing body of evidence indicates that center symmetry is
intimately related also with another aspect of the physics of fermions
that changes radically as the system transitions to the high
temperature phase, namely the localization properties of the low-lying
eigenmodes of the Dirac operator (see Ref.~\cite{Giordano:2021qav} for
a recent review).  It has been shown in a rather large variety of
gauge theories, including QCD, that as the theory crosses over from
the confined to the deconfined phase, the low Dirac modes turn from
delocalized to localized, up to a ``mobility edge'' $\lambda_c$ in the
spectrum, above which they are again
delocalized~\cite{GarciaGarcia:2005vj,
  GarciaGarcia:2006gr,Kovacs:2009zj,Kovacs:2010wx,Kovacs:2012zq,
  Giordano:2013taa,Nishigaki:2013uya,Giordano:2014qna,
  Ujfalusi:2015nha,Giordano:2016nuu,Cossu:2016scb,Holicki:2018sms,
  Kovacs:2017uiz,Giordano:2019pvc,Vig:2020pgq,Bonati:2020lal,
  Baranka:2021san}. The strong connection observed between
localization and deconfinement is clarified by the mechanism provided
by the ``sea/islands picture'' of localization proposed in
Refs.~\cite{Bruckmann:2011cc,Giordano:2015vla,Giordano:2016cjs,
  Giordano:2016vhx}. This picture relates localized modes with local
fluctuations of the Polyakov loop away from its ordered value (i.e.,
1) in the high-temperature phase, which provide ``energetically''
favorable locations for the fermions. Numerical support for this
picture has been obtained in various
cases~\cite{Cossu:2016scb,Holicki:2018sms,Baranka:2021san,Bruckmann:2011cc}.

Since in QCD the transition is an analytic crossover, the statement
that localization of the low modes and deconfinement happen together
can only be of qualitative nature. Localization in the presence of a
sharp transition has been mostly investigated in pure gauge theories,
selecting the ``physical'' center sector (i.e., real positive
expectation value of the Polyakov loop) in the spontaneously broken
phase. In these cases localization and deconfinement have been shown
to coincide within numerical
uncertainties~\cite{Kovacs:2017uiz,Giordano:2019pvc,
  Vig:2020pgq,Bonati:2020lal,Baranka:2021san}. The only study with
dynamical fermions and a genuine phase transition is that of
Ref.~\cite{Giordano:2016nuu}, concerning the SU(3) theory in the
presence of unimproved staggered fermions on coarse lattices, where
again localization was found to appear exactly at the
transition. However, while this model is well defined as a statistical
system on the lattice, the transition is known to be only a lattice
artifact~\cite{Karsch:2001nf,deForcrand:2003vyj,deForcrand:2008vr}.

An interesting and as yet unexplored scenario is that of a genuine
phase transition in the presence of dynamical fermions that survives
the continuum limit. The Roberge-Weiss transition precisely provides
such a scenario. In fact, the imaginary chemical potential effectively
modifies the ``twist'' imposed on the Dirac modes by the antiperiodic
boundary conditions, and favors configurations where the Polyakov
loops most effectively ``neutralize'' it. For gauge group SU(3) and
$\hat{\mu}_I=\pi$, these correspond to the two complex sectors
$e^{\pm i\frac{2\pi}{3}}$, which leaves an exact $\mathbb{Z}_2$ center
symmetry that can break down spontaneously. This happens at
$T_{\rm RW}$, where the system undergoes a second-order phase
transition to a deconfined phase, where either of the two complex
sectors can be selected, and the local Polyakov loops prefer to align
to either $e^{i\frac{2\pi}{3}}$ or $e^{- i\frac{2\pi}{3}}$. This opens
a pseudogap of low spectral density in the Dirac spectrum, equal to
the effective Matsubara frequency $\omega_{\rm RW} = (2\pi/ 3)
T$. According to the sea/islands picture (suitably adapted to the case
of nonzero imaginary chemical potential), this pseudogap can be
populated by localized modes living on the fluctuations of the
Polyakov loop away from the ordered value. One then expects the
low-lying Dirac modes to turn from delocalized to localized as
$T_{\rm RW}$ is crossed. Confirming this scenario would lend more
support to the conjectured strong connection between localization and
deconfinement.

In this paper we study this scenario by means of numerical simulations
on the lattice. In particular, we consider $N_f = 2+1$ QCD with
physical quark masses, discretized via improved staggered fermions,
with a degenerate imaginary chemical potential coupled to all quark
flavors, i.e., we consider a purely baryonic imaginary chemical
potential.  We then determine the localization temperature, at which
the lowest modes turn from delocalized to localized, at finite spacing
for different values $N_t = 4, 6, 8$ of the temporal extension of the
lattice. This is compared to the critical temperature of the
Roberge-Weiss transition at the same $N_t$, as well as to its
continuum extrapolation, obtained in Ref.~\cite{Bonati:2016pwz}
adopting the same discretization used in this study.

The paper is organized as follows. In Section~\ref{sec:numsetup} we
provide details about the system discretization and the numerical
algorithms adopted in our investigation; in Section~\ref{sec:numres}
we discuss our numerical results; finally, in Section~\ref{sec:concl}
we draw our conclusions. Systematic effects on the determination of
the mobility edge near the Roberge-Weiss transition are discussed in
the Appendix.

\section{Numerical setup}
\label{sec:numsetup} 

\subsection{Dirac spectrum and localization}
\label{sec:numsetup_local}

The staggered operator at nonzero $\hat{\mu}_I=\mu_I/T$ reads
\begin{equation}
  \label{eq:muI1}
  \begin{aligned}
    (D_{\rm stag}(\hat{\mu}_I))_{xy} =
    {\textstyle\frac{1}{2}}\sum_\alpha&\eta_\alpha(x)(
    e^{i\frac{\hat{\mu}_I}{N_t} \delta_{\alpha 4}} U_\alpha(x)
    \delta_{x+\hat{\alpha}\,y} \\ - & e^{-i\frac{\hat{\mu}_I}{N_t}
      \delta_{\alpha 4}} U_\alpha^\dag(x-\hat{\alpha})
    \delta_{x-\hat{\alpha}\,y})\,,
  \end{aligned}
\end{equation}
where $U_\alpha(x)\in {\rm SU}(3)$, $\alpha=1,\ldots,4$ are the link
variables and $\eta_\alpha(x)$ are the usual staggered phases.
Periodic boundary conditions in the spatial directions, and
antiperiodic boundary conditions in the temporal direction, are
understood. The operator $D_{\rm stag}(\hat{\mu}_I)$ is anti-Hermitian
and has the chiral property $\{\eta_5,D_{\rm stag}(\hat{\mu}_I)\}=0$,
where $(\eta_5)_{xy}=\eta_5(x)\delta_{xy}$ with
$\eta_5(x)=(-1)^{\sum_{\alpha=1}^4 x_\alpha}$.  The spectrum of
$D_{\rm stag}(\hat{\mu}_I)$ is purely imaginary due to
anti-Hermiticity,
\begin{equation}
  \label{eq:stagspec1}
  D_{\rm stag}(\hat{\mu}_I) \psi_n = i\lambda_n \psi_n\,, 
\end{equation}
with $\lambda_n\in \mathbb{R}$, and furthermore symmetric with respect
to zero thanks to the chiral property, since this implies
$D_{\rm stag}(\hat{\mu}_I) \eta_5\psi_n = -i\lambda_n \eta_5\psi_n$.  This
implies in particular that $\det [D_{\rm stag}(\hat{\mu}_I) + m]$ is real
and positive.

The localization properties of the eigenmodes are determined by the
large-volume scaling of the Inverse Participation Ratio (IPR),
averaged over gauge configurations. The IPR is defined as:
\begin{equation}
  \label{eq:stagspec2}
  {\rm IPR}_n = \textstyle\sum_x \left(\sum_c |(\psi_n(x))_c |^2 \right)^2\,,
\end{equation}
where $c=1,2,3$ is the color index. For modes effectively occupying a
region of size $V_{\rm eff}$, one finds
$\sum_c |(\psi_n(x))_c |^2 \sim 1/V_{\rm eff}$ and so qualitatively
$\langle {\rm IPR}_n\rangle\sim V_{\rm eff}/V_{\rm eff}^2 = 1/V_{\rm
  eff}$.  Delocalized modes are extended all over the space, i.e.,
$V_{\rm eff}\sim V$, and so $\langle {\rm IPR}_n\rangle\sim 1/V\to 0$
as $V\to\infty$, while for localized modes $V_{\rm eff} \sim V_0$ is
finite and $V$-independent, so that one expects ${\rm IPR}_n$ to
remain constant as $V\to\infty$. Since $\psi_n$ and $\eta_5\psi_n$
have the same IPR, it suffices to focus on $\lambda_n\ge 0$ only.

The localization properties of the eigenmodes $\psi_n$ are most easily
studied by exploiting their connection with the statistical properties
of the spectrum~\cite{altshuler1986repulsion}. In fact, localized
modes are expected to fluctuate independently and so the corresponding
eigenvalues are expected to obey Poisson statistics. For delocalized
modes the corresponding eigenvalues are expected instead to obey the
same statistics as the appropriate Gaussian ensemble of Random Matrix
Theory (RMT)~\cite{mehta2004random}, once model-dependent features are
removed by the unfolding procedure. For $D_{\rm stag}$ and SU(3) gauge
fields, the right ensemble is the Gaussian unitary ensemble
(GUE)~\cite{Verbaarschot:2000dy}. The unfolded spectrum is defined by
the mapping
\begin{equation}
  \label{eq:stagspec3}
  x_n = \int^{\lambda_n} d\lambda'\, \rho(\lambda')\,, 
\end{equation}
where $\rho(\lambda) = \langle \sum_n\delta(\lambda-\lambda_n)\rangle$
is the spectral density. In practice, $x$ is the expected ranking of
an eigenvalue equal to $\lambda$ if this is found on a configuration.
Convenient observables are obtained from the probability distribution
$p(s;\lambda)$ of the unfolded level spacings $s_n=x_{n+1}-x_n$,
computed locally in the spectrum,
\begin{equation}
  \label{eq:stagspec5}
  p(s;\lambda) = \frac{\left\langle {\textstyle
        \sum_n}\delta(\lambda-\lambda_n)\delta(s-s_n)\right\rangle}{\left\langle
      {\textstyle \sum_n}\delta(\lambda-\lambda_n)\right\rangle}\,. 
\end{equation}
For localized, independently fluctuating modes one expects
$p(s;\lambda)$ to be that corresponding to Poisson statistics,
\begin{equation}
  \label{eq:stagspec6}
  p_{\rm Poisson}(s) = e^{-s}\,,
\end{equation}
while for delocalized modes $p(s;\lambda)$ should be equal to that of
the GUE, which is well approximated by the so-called Wigner surmise,
\begin{equation}
  \label{eq:stagspec7}
  p_\rmt(s) = \frac{32}{\pi^2} s^2 e^{-\frac{4}{\pi}s^2}\,.
\end{equation}
The transition from localized to delocalized modes can be monitored by
looking at how the features of $p(s;\lambda)$ change across the
spectrum. To this end, in this paper we have used the integrated
probability distribution~\cite{Shklovskii:1993zz},
\begin{equation}
  \label{eq:stagspec8}
  I_{s_0}(\lambda) = \int_0^{s_0}ds \,p(s;\lambda)\,,
\end{equation}
where $s_0\simeq 0.508$ is chosen to maximize the difference between
Poisson and RMT-type statistics. For these statistics one finds
$I_{s_0,\, {\rm Poisson}}\simeq 0.398$ and
$I_{s_0,\, \rmt}\simeq 0.117$.  Moreover, the critical value at the
mobility edge is known for this observable,
$I_{s_0,\, {\rm crit}} = I_{s_0}(\lambda_c) =
0.1966(25)$~\cite{Giordano:2013taa}, and is expected to be universal.
This can be used to identify the position of the mobility edge
$\lambda_c$ with good precision without the need for a finite-size
scaling study, by simply looking at the crossing point of some
interpolation of the numerical data and $I_{s_0,\, {\rm crit}}$.

\subsection{Simulation details}
\label{sec:techdet}

We studied $N_f=2+1$ QCD on $N_s^3\times N_t$ hypercubic lattices
using 2-stout improved~\cite{Morningstar:2003gk} rooted staggered
fermions with physical quark masses, at finite temperature
$T=1/(aN_t)$ and in the presence of an imaginary chemical potential
$\mu_I$.  Gauge configurations were generated using a Rational Hybrid
Monte-Carlo algorithm running on GPUs~\cite{Bonati:2018wqj}.  Details
about the implementation can be found in Ref.~\cite{Bonati:2016pwz}.
For what follows it is useful to remember that the bare parameters,
including the quark masses, are tuned so as to stay on a line of
constant physics while changing the ultraviolet cut-off of the theory,
following the determination reported in
Refs.~\cite{Aoki:2009sc,Borsanyi:2010cj,Borsanyi:2013bia}.

To study localization above the Roberge-Weiss point, we set
$\hat{\mu}_I=\pi$ and performed a scan in temperature at
$T>T_{\rm RW}$. We then computed the low modes of the staggered Dirac
operator numerically using the ARPACK
library~\cite{lehoucq1998arpack}.  We started using $N_t=4$ and doing
a preliminary check for finite-volume effects, comparing the mobility
edges obtained at a given temperature according to the procedure
discussed above on lattices of increasing spatial dimension (this
procedure is described in more detail in Sec.~\ref{sec:numres}). In
particular, we used $N_s=24,32,40$ at $T=394\,{\rm MeV}$ and
$N_s=24,32$ at $T=197\,{\rm MeV}$, finding in both cases compatible
results for $\lambda_c$ from the various volumes. This means that an
aspect ratio $r_{st}\equiv N_{s}/N_{t} = 24/4=6$ is expected to
already reproduce well the thermodynamic limit. We then used also
$N_{t} = 6,8$ to check for finite-spacing effects. Compatibly with the
numerical effort, we used $r_{st}=8$ for $N_{t}=4$ and $r_{st}=6$ for
$N_{t} = 6,8$.

At $\hat{\mu}_I=\pi$ the dynamics favors equally the two complex
center sectors $z=e^{\pm i\frac{2\pi}{3}}$ over the real sector
$z=1$. In the thermodynamic limit (taken in the presence of a suitable
infinitesimal perturbation breaking the residual $\mathbb{Z}_2$
symmetry) only one of the two complex sectors survives. In a finite
volume, instead, the system tunnels between the two complex sectors
(and with a much smaller probability, it can also tunnel to the real
sector). For what concerns the staggered spectrum, however, one need
not treat them separately to obtain the correct result in the
thermodynamic limit. In fact, the spectrum of
$D_{\rm stag}(\hat{\mu}_I=\pi)=D_{\rm stag}^{\rm PBC}$ (i.e., with
periodic boundary condition in the temporal direction) on
configuration $U$ belonging to the center sector $z$ is equal to the
spectrum on configuration $U^*$ belonging to the center sector $z^*$.
If $\lambda,\psi$ is an eigenpair on configuration $U$, one has
\begin{equation}
  \label{eq:ccinv}
    D_{\rm stag}^{\rm PBC}[U^*]\eta_5\psi^* = \left(-\eta_5D_{\rm
        stag}^{\rm PBC}[U]\psi\right)^*  =
    i\lambda\eta_5\psi^*\,.
\end{equation}
In particular, this means that $U$ and $U^*$ have the same Boltzmann
weight (in the absence of $\mathbb{Z}_2$-breaking perturbations).  Any
configuration belonging to the sector $e^{- i\frac{2\pi}{3}}$
appearing in the simulation history can then be treated effectively as
just another configuration in the sector $e^{+ i\frac{2\pi}{3}}$ (and
viceversa), with the same spectrum, and with the correct weight if one
restricted the configuration space to a single center sector (up to
finite-size effects due to contaminations from the real sector). We
have then not imposed any restriction on the configurations to be
analyzed, and included them all in the analysis. Since we observed no
tunnelling to the real sector in our simulation histories, the
corresponding finite-size effects are absent.

\section{Numerical results}
\label{sec:numres}

\subsection{Determination of the mobility edge}
\label{sec:mobedge}

To unfold the spectrum we collected all the eigenvalues in the
ensemble of configurations obtained for a given lattice setup, ranked
them by magnitude, and replaced them by their rank divided by the
number of configurations. We then divided the spectrum in bins, and
computed $I_{s_0}(\lambda)$ in each bin separately, assigning the
result to the central point of the bin. To ensure that bins are
sufficiently small to reliably capture the local behavior of
$I_{s_0}(\lambda)$, we have computed the average unfolded spacing
$\langle s \rangle_\lambda$ in each bin.  This should equal 1, and we
have checked that this is the case in the relevant spectral
regions.\footnote{The relation $\langle s\rangle_\lambda = 1$ follows
  from the fact that for infinite statistics and in the large-volume
  limit the average level spacing
  $\langle\Delta\lambda\rangle_\lambda$ in an infinitesimal spectral
  region around $\lambda$ equals $1/\rho(\lambda)$, and this is
  identically 1 for the unfolded spectrum by construction. For finite
  statistics and volume one has necessarily to use sufficiently large
  finite bins in order to collect sufficiently many eigenvalues, and
  in regions where $\rho(\lambda)$ is small the bin size may be
  comparable or even exceed the scale over which $\rho(\lambda)$
  varies appreciably. This leads to
  $\rho(\lambda)\langle\Delta\lambda\rangle_\lambda \neq 1$ and in
  turn to $\langle s\rangle_\lambda\neq 1$ in that
  region~\cite{Giordano:2019pvc}. This happens in the lowest part of
  the staggered spectrum at high temperature where the spectral
  density is small. This region is problematic also due to the effects
  of the approximate taste symmetry of staggered fermions at finite
  lattice spacing, which distorts the spectral statistics from
  Poissonian~\cite{Kovacs:2011km,Kovacs:2012zq}. Moreover, since we
  computed a limited and fixed number of eigenvalues for each
  configuration, cut-off effects lead to
  \protect{$\langle s \rangle_\lambda \neq 1$} also at the highest end
  of the spectral region being explored. However, both these spectral
  regions are irrelevant to our analysis and were discarded.}

\begin{figure}[t]
\includegraphics[width=0.9\columnwidth, clip]{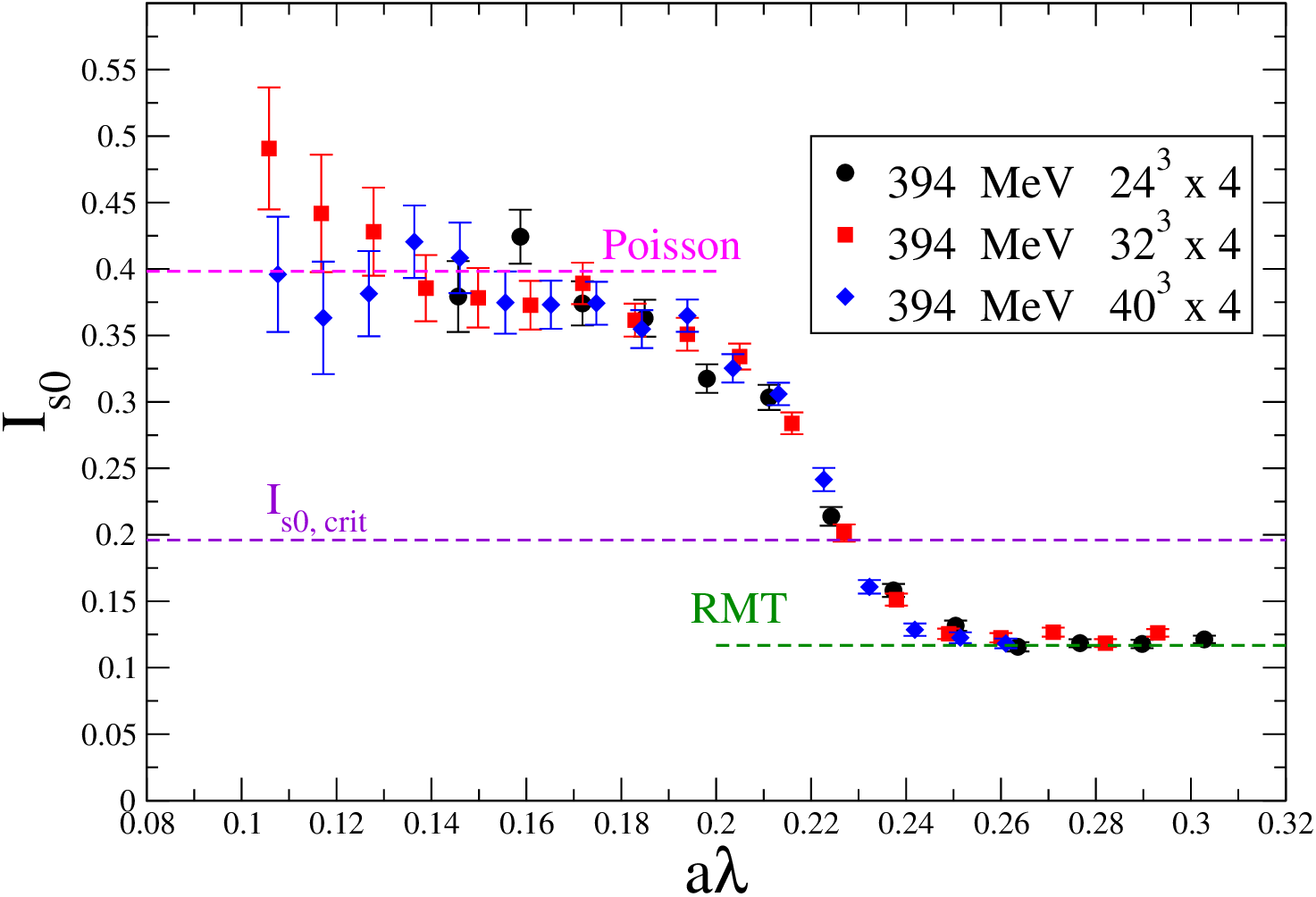}
\caption{Integrated probability distribution of the unfolded level
  spacings, computed locally in the spectrum, at $T=394\,{\rm MeV}$ on
  $N_s^3\times 4$ lattices. Horizontal lines correspond to the
  expectation for Poisson statistics (corresponding to localized
  modes), RMT statistics (corresponding to delocalized modes), and
  critical statistics. An Anderson transition in the Dirac spectrum is
  found at the mobility edge where the curve intersects the critical
  value $I_{s_0,{\rm crit}}$.}
\label{fig:AT}
\end{figure}
\begin{figure}[t]
\includegraphics[width=0.9\columnwidth, clip]{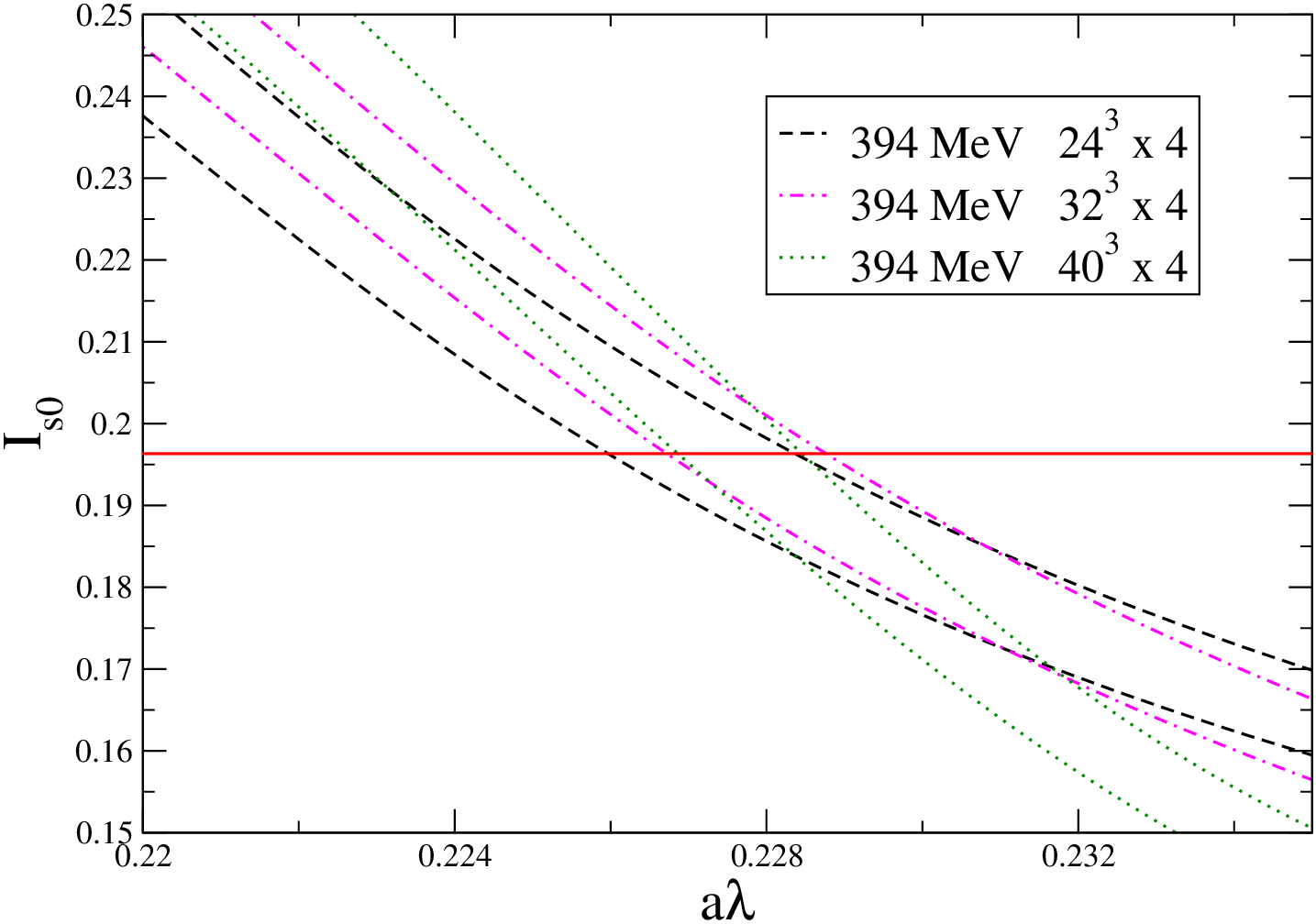}
\caption{Spline interpolations of $I_{s_{0}}\pm \delta I_{s_{0}}$ used
  for the determination of the mobility edge. The horizontal line
  corresponds to $I_{s_0,{\rm crit}}$, and its crossing points with
  the spline interpolations determine the error band for
  $\lambda_c$. Different spatial volumes yield consistent results.}
\label{fig:splines}
\end{figure}

Results of this procedure are shown in Fig.~\ref{fig:AT}, for
$T=394\,{\rm MeV}$, $N_t=4$, and the three different spatial
volumes. The presence of localized modes at the low end of the
spectrum is signalled by $I_{s_0}\approx I_{s_0,\,{\rm Poisson}}$, while
in the bulk $I_{s_0}\approx I_{s_0,\,\rmt}$ indicates that modes are
delocalized.  At the mobility edge $\lambda_c$ where the Anderson
transition from localized to delocalized modes takes place, $I_{s_0}$
takes the critical value $I_{s_0}= I_{s_0,\, {\rm crit}}$.

To determine $\lambda_c$, we constructed two cubic spline
interpolations of $I_{s_{0}}\pm\delta I_{s_{0}}$, where
$\delta I_{s_{0}}$ is the statistical error on $I_{s_{0}}$, and looked
for the crossing points of the spline interpolations with
$I_{s_0,\, {\rm crit}}$. We then determined $\lambda_c$ as the average
of the crossing points, with an associated error equal to the
semi-dispersion. We also checked that changing the order of the spline
interpolation leads to negligible effects on the determination of
$\lambda_c$. The procedure is illustrated for $T=394\,{\rm MeV}$,
$N_t=4$, and the three different volumes in Fig.~\ref{fig:splines}.
Strictly speaking, the mobility edge is the point in the spectrum
where $I_{s_0}$ is volume-independent. Figure \ref{fig:splines} shows
that our procedure yields consistent values for the three volumes,
thus providing an accurate estimate for $\lambda_c$, and that
statistical fluctuations dominate over the finite-size effects.

\subsection{Localization properties at the Roberge-Weiss phase
  transition}

\begin{figure}[t]
  \includegraphics[width=0.9\columnwidth, clip]{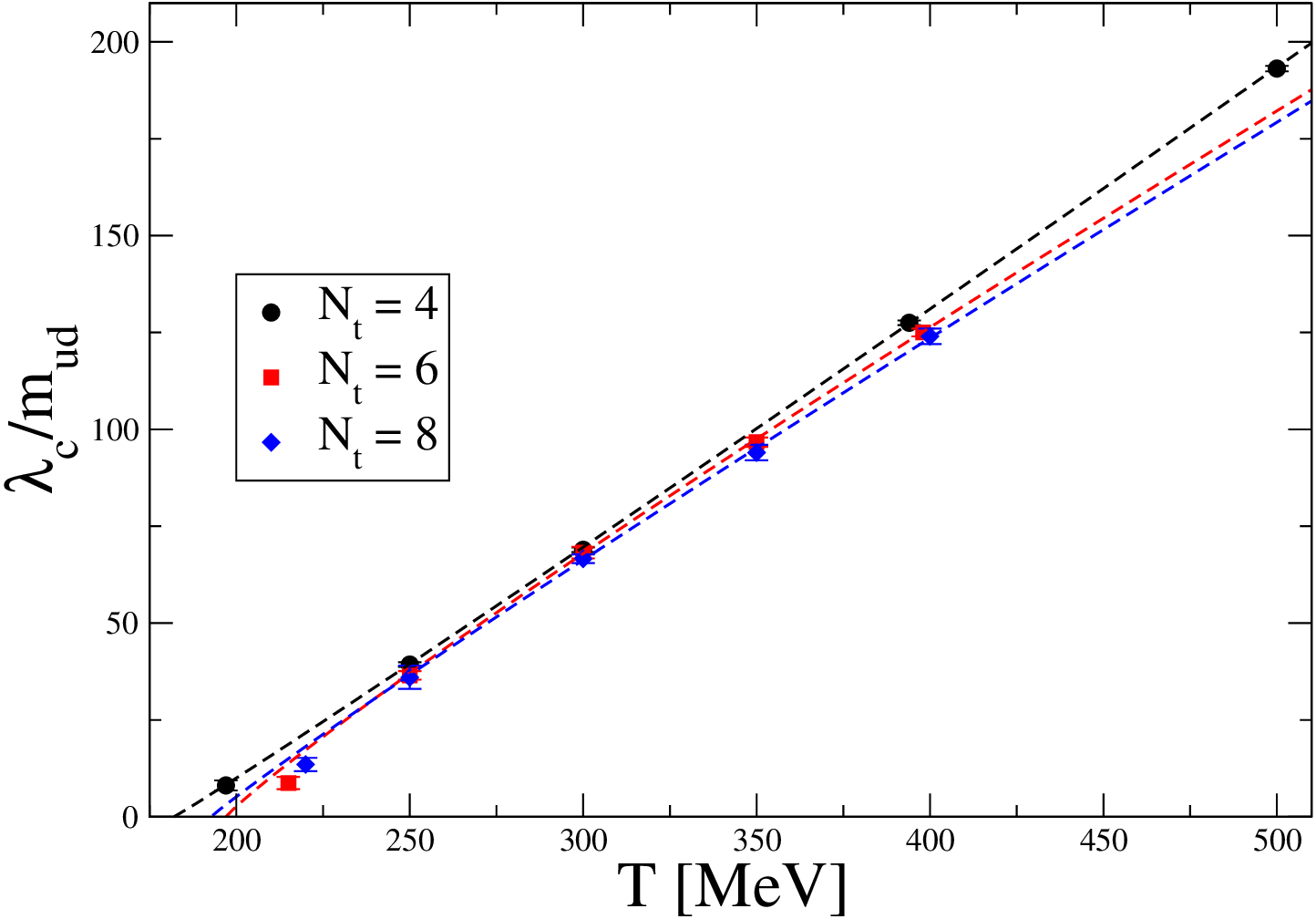}
  \caption{Temperature dependence of the renormalized mobility edge.
    Dashed lines are best fits to the data at fixed $N_t$ with the
    functional form Eq.~\eqref{eq:fitform}, excluding the lowest
    temperature from each set.}
\label{fig:lc_m}
\end{figure}

\begin{table}[t]
  \centering
  \begin{tabular}{c|ccc}
    $N_t$  & $A$ & $B$ & $T_{\rm loc}\, [{\rm MeV}]$ \\ \hline
    4 &  $0.51(1)$ & $1.0(1)$ & $183(4)$         \\
    6 &  $0.98(3)$ & $0.91(5)$ & $198(6)$        \\
    8 &  $0.8(4)$  & $0.94(2)$ & $193(10)$       
  \end{tabular}
  \caption{Coefficients of the best fit of the renormalized mobility
    edge to the functional form
    $\tilde{\lambda}_c(T) = A(T-T_{\rm loc})^B$.}
  \label{tab:fitresults}
\end{table}

The results for the mobility edge, determined as discussed above, are
collected in Fig.~\ref{fig:lc_m}. There we show the
renormalization-group-invariant ratio
$\tilde{\lambda}_c=\frac{\lambda_c}{\mlight}$,\footnote{Roughly
  speaking, the eigenvalues of the Dirac operator renormalize
  multiplicatively with the same renormalization constant as the quark
  mass~\cite{DelDebbio:2005qa,Giusti:2008vb}, and so is expected to do
  the mobility edge~\cite{Kovacs:2012zq}.  A proof that this is
  actually the case will be presented
  elsewhere~\cite{giordano_inprep}.  Therefore, taking its ratio with
  the quark mass, which is tuned so as to stay on a line of constant
  physics, returns a renormalization-group invariant quantity.} with
$\mlight$ the bare light-quark mass, for all the temperatures and
lattice spacings used in this work. The dependence on the lattice
spacing is indeed mild, with $\tilde{\lambda}_c(T,N_t)$ depending
little on $N_t$.

It is evident that the mobility edge tends to vanish as $T$ decreases
toward $T_{\rm RW}$.  As the Roberge-Weiss transition, at least for
$N_f = 2+1$ QCD with physical quark masses and close enough to the
continuum limit, is a continuous
transition~\cite{Bonati:2016pwz,Bonati:2018fvg}, we also expect
$\tilde{\lambda}_c$ to vanish continuously. To leading order, we then
expect
\begin{equation}
  \label{eq:fitform}
\tilde{\lambda}_c(T,N_t) = A(N_t)\left[T-T_{\rm loc}(N_t)\right]^{B(N_t)}\,.  
\end{equation}
For each $N_t$ we separately determined the localization temperature
$T_{\rm loc}(N_t)$ where the mobility edge vanishes, by fitting the
data to the functional form Eq.~\eqref{eq:fitform}. In doing so, we
have excluded the lowest temperature, closest to the transition, from
each set. As we show in the Appendix, the determination of $\lambda_c$
is still affected by systematic effects at the lowest temperature for
the $N_t=6,8$ ensembles, which could be due to finite size effects
becoming more visible as the continuum limit is approached (see the
Appendix for more details).  Since this data point affects the outcome
of the fit and may distort the result for $T_{\rm loc}(N_t)$, we have
preferred to use only the more reliable results obtained at
temperatures fairly and well above $T_{\rm RW}$.
\begin{figure}[t!]
  \includegraphics[width=0.9\columnwidth, clip]{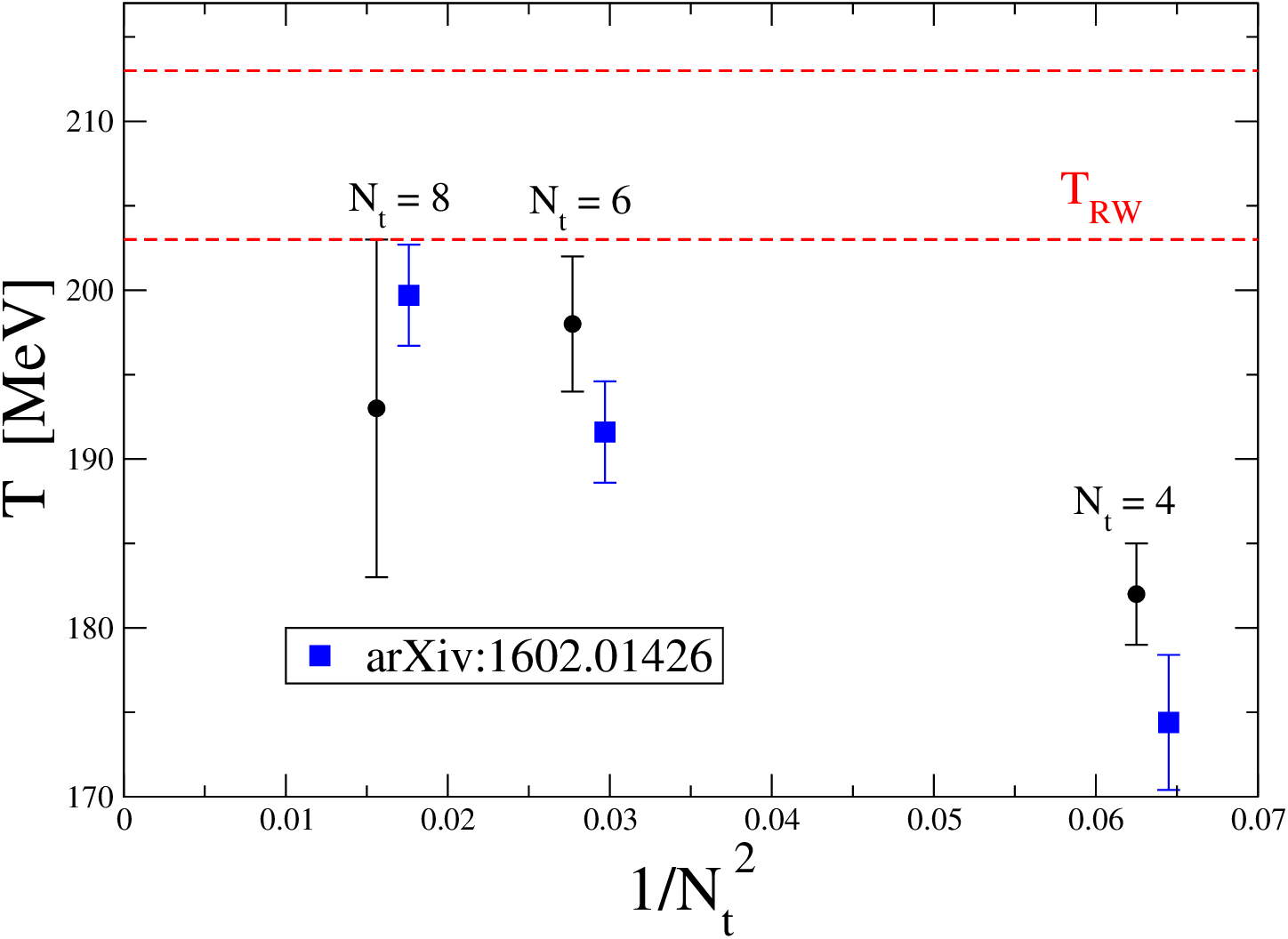}
  \caption{Localization temperature $T_{\rm loc}(N_t)$, where
    $\tilde{\lambda}_c$ vanishes, for the various $N_t$s. For
    comparison, we show also the Roberge-Weiss temperatures
    $T_{\rm RW}(N_t)$ at finite spacing (slightly shifted
    horizontally) and the error band corresponding to its continuum
    extrapolation, obtained in Ref.~\cite{Bonati:2016pwz}.}
\label{fig:contlim}
\end{figure}

Results for $T_{\rm loc}(N_t)$ are shown in Fig.~\ref{fig:contlim},
together with the critical temperatures $T_{\rm RW}(N_t)$ at finite
spacing and the band corresponding to the continuum-extrapolated
result $T_{\rm RW}=208(5)\,{\rm MeV}$ for the Roberge-Weiss
temperature, obtained in Ref.~\cite{Bonati:2016pwz}.  We find that
$T_{\rm loc}(N_t)$ is compatible with $T_{\rm RW}(N_t)$ for all
$N_t$s. A continuum extrapolation via a fit linear in $1/N_t^2$ gives
$T_{\rm loc}=204(7)\,{\rm MeV}$, in good agreement with $T_{\rm RW}$.

\section{Conclusions} 
\label{sec:concl}

We have studied the localization properties of the low-lying modes of
the staggered operator at imaginary chemical potential $\mu_I/T = \pi$
in $N_f = 2+1$ QCD above the Roberge-Weiss temperature $T_{\rm RW}$,
by means of numerical lattice simulations with rooted staggered
fermions at physical quark masses. We found that the low modes are
localized up to a temperature- and spacing-dependent mobility edge
$\lambda_c$, that is extrapolated to vanish at $T_{\rm loc}(N_t)$. For
the renormalized mobility edge $\lambda_c/\mlight$, and sufficiently
above $T_{\rm RW}$, we observed only a mild dependence on $N_t$, as
expected. For the localization temperatures $T_{\rm loc}(N_t)$ where
the mobility edge vanishes for the various $N_t$s we obtained values
in agreement with the determination of the critical temperatures
$T_{\rm RW}(N_t)$ of Ref.~\cite{Bonati:2016pwz}. The same is true for
their continuum extrapolations. This supports the expectation that
localized modes appear precisely at the deconfinement transition of a
gauge theory, when such a transition is sharp. In particular, this is
the first case when the close connection between localization of the
low Dirac modes and deconfinement is demonstrated for a genuine
deconfinement transition in the presence of dynamical fermions that
survives the continuum limit.

\appendix

\section{Systematic effects near $T_{\rm RW}$}
\label{sec:app}

\begin{table}[tbh!]
  \centering
  \begin{tabular}{c|c|c|c}
    \multirow{2}{*}{$N_t$} & \multirow{2}{*}{$T_{\rm min}\,[\rm {MeV}]$}
    & $T_{\rm loc}\,[{\rm MeV}]$
    & $T_{\rm loc}\,[{\rm MeV}]$  \\
                           &  &  ($T_{\rm min}$ excluded) & ($T_{\rm
                                                            min}$ included) \\
    \hline \hline
    4 & $197$ & $183(4)$  & $183(2)$ \\
    6 & $215$  & $198(6)$  & $206(2)$ \\
    8 & $220$  & $193(10)$  & $205(3)$
  \end{tabular}
  \caption{$T_{\rm loc}$ obtained from fits excluding or including the
    lowest temperature $T_{\rm min}$ for the various $N_t$.}
  \label{tab:fitres}
\end{table}

To further check for finite-size and other systematic effects near the
transition, for each $N_t$ we compared the results of fits to
$\tilde{\lambda}_c(T,N_t)$ performed including or excluding the lowest
temperature from each set. The temperature closest to the
Roberge-Weiss transition is in fact the one affected the most by
finite-size effects due to the larger correlation length.  In
particular, taste-violating effects become milder as the lattice
becomes finer, leading to the formation of multiplets of low modes
that distort the spectral statistics. This can affect our
determination of the mobility edge when this is close to zero,
especially on our finer $N_t=6,8$ ensembles.

\begin{figure}[ht!]

  \includegraphics[width=0.9\columnwidth, clip]{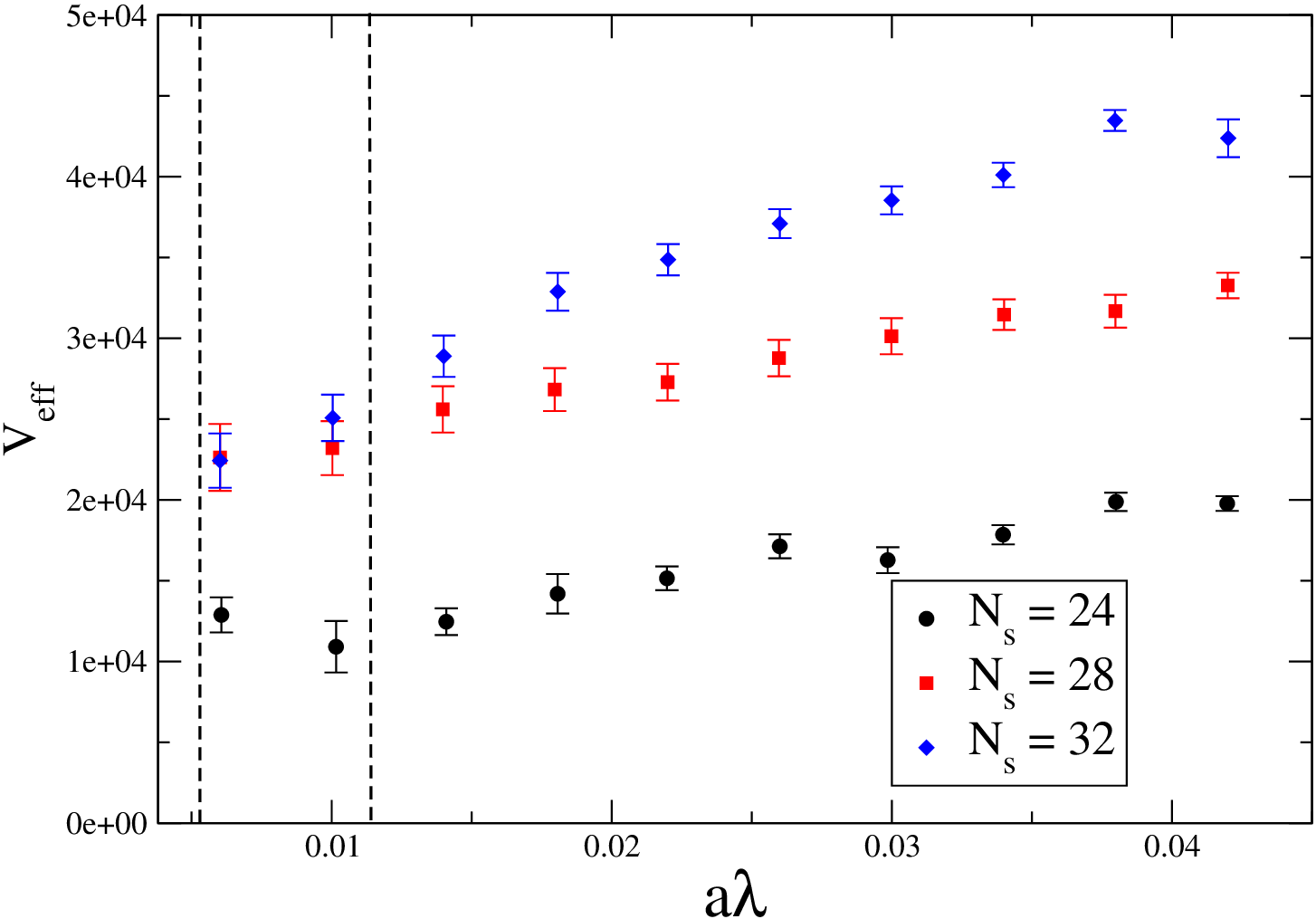}
  \caption{Size $V_{\rm eff} ={\rm IPR}^{-1}$ of low modes on $N_t=6$
    lattices at $T=200\,{\rm MeV}$ for $V=N_s^3=24^3,28^3,32^3$. For
    the lowest modes $a\lambda\lesssim 0.01$ this is approximately the
    same for the two largest volumes, indicating that they are
    localized.}
\label{fig:nt6T200}
\end{figure}

It is already evident from Fig.~\eqref{fig:lc_m} that including the
lowest temperature in a fit of the form Eq.~\eqref{eq:fitform} will
alter the result for $T_{\rm loc}(N_t=6,8)$. The results are reported
in Tab.~\ref{tab:fitres}. For $N_t=4$ the two results are compatible:
this is not surprising, given the relative coarseness of the lattice
and the larger aspect ratio. For $N_t=6$ and $N_t=8$, instead, the
difference is substantial, and around 4\% and 6\%, respectively.

To show explicitly that the determination of $T_{\rm loc}(N_t=6)$ is
inaccurate if the lowest temperature $T=215\,{\rm MeV}$ is included,
we have done a short run at $T=200\, {\rm MeV}$ on three volumes,
$N_s=24,28,32$. Below $a\lambda\simeq 0.01$ the size of the modes,
$V_{\rm eff}={\rm IPR}^{-1}$, is compatible within 1$\sigma$ for the
$N_s=28,32$ ensembles (see Fig.~\ref{fig:nt6T200}).  This is a clear
indication that the lowest modes are localized, and so that
necessarily $T_{\rm loc}(N_t=6) < 200 \,{\rm MeV}$. We take this as an
indication that our sample of $N_t=6$ configurations at
$T=215\,{\rm MeV}$ is affected by strong systematic effects, which are
likely a combination of finite-size effects and limited statistics.
Instead, the value obtained exluding the lowest temperature is
compatible with this finding.

\vspace{0.5cm}
\paragraph*{Acknowledgments}

We thank G.~Clemente and F.~Sanfilippo for helping implement the
diagonalization code.  Numerical simulations have been performed on
the MARCONI and MARCONI100 machines at CINECA, based on the agreement
between INFN and CINECA (under projects INF20 npqcd, INF21 npqcd).
M.~G.\ is partially supported by the NKFIH grant KKP-126769.

\bibliographystyle{apsrev4-2}
\bibliography{biblio_merlo}

\end{document}